\documentstyle[proceedings,psfig]{crckapb} 

\begin{opening}
\title{Type Ia Supernovae as Extragalactic Distance Indicators}

\author{David Branch}
\author{Peter Nugent}
\author{Adam Fisher}
\institute{University of Oklahoma\\
           Department of Physics and Astronomy\\
           Norman, OK, 73019, USA}

\end{opening}

\begin{document}

\begin{abstract}
Because Type Ia supernovae (SNe Ia) are not {\sl perfect} standard
candles, it is important to be able to use distance--independent
observables (DIOs) to define subsets of SNe Ia that are ``nearly
standard candles'' or to correct SN Ia absolute magnitudes to make
them nearly homogeneous (``standardized candles'').  This is not
crucial for the measurement of $H_0$, but it is for the measurement of
$q_0$ and of parent--galaxy peculiar velocities.  We discuss the use
of various photometric and spectroscopic SN Ia DIOs, and a
parent--galaxy DIO, for this purpose.  We also discuss the status of
the absolute--magnitude calibration of SNe Ia. We find that SNe Ia,
whether calibrated by means of (1) Cepheids in their parent galaxies,
(2) fitting their optical--ultraviolet spectra with detailed non-LTE
model atmosphere calculations, or (3) by considering that the light
curve is powered by the decay of radioactive $^{56}$Ni, firmly
indicate that the value of $H_0$ is low, less than or about 60 $\rm
km\ s^{-1}\ Mpc^{-1}$. Some issues regarding the determination of
$q_0$ by means of SNe Ia are discussed briefly.  Finally, we
conjecture that even if $q_0$=0.5, there probably is no cosmic age
problem.
\end{abstract}

\section{Introduction}
Interest in using Type Ia supernovae (SNe~Ia) as extragalactic
distance indicators has increased in recent years, as important new
data have been obtained and as our understanding of the basic physics
of SNe~Ia has improved.  SNe~Ia are bright enough to be observed to
cosmologically significant distances, the distribution of their
absolute magnitudes at maximum light is sharply peaked, and the
absolute magnitudes can be calibrated in several independent ways.

SNe~Ia are not {\sl perfect} standard candles.  In principle, this
poses no obstacle for the determination of $H_0$, because given
appropriate observational data for an individual SN~Ia out in the
Hubble flow, physical modeling can be applied to determine the
absolute magnitude, the distance, and the value of $H_0$, without any
appeal to absolute-magnitude homogeneity.  In practice, however, most
of the work that has been done so far on $H_0$ from SNe~Ia has
involved some matching of the absolute magnitudes of SNe~Ia out in the
Hubble flow to those of nearer SNe~Ia whose absolute magnitudes have
been calibrated either by physical modeling or by means of Cepheid
variables in their parent galaxies.  Matching of remote and nearer
SNe~Ia will continue to be required for applications that in practice
must remain statistical in nature - determining the value of $q_0$ and
measuring galaxy peculiar velocities.

It can be helpful to keep these matters in quantitative perspective.
The difference between $H_0=85$ and $H_0=55$, expressed in magnitudes,
is $5 Log(85/55) = 0.95$ mag.  Thus, making a useful measurement of
$H_0$ from SNe~Ia is {\sl not} a delicate matter, because it is easy
to define subsamples of SNe~Ia whose absolute-magnitude dispersions are
much less than that. The issue is the absolute-magnitude calibration.
We will argue below that SNe~Ia firmly indicate that the value of
$H_0$ is low, less than or about 60 $\rm km\ s^{-1}\ Mpc^{-1}$.

Using high-redshift SNe~Ia to determine the value of $q_0$ {\sl is} a
delicate matter.  The absolute-magnitude calibration doesn't enter,
but at $z \simeq 0.5$ the difference between $q_0=0.5$ and $q_0=0.1$
is only about 0.2 magnitudes.  Therefore measuring $q_0$ to an
accuracy of 0.1 requires matching the magnitudes of remote SNe~Ia,
relative to nearer ones, to within about 0.05 magnitudes.  Obtaining
this kind of accuracy requires a statistical approach, i.e., beating
down the uncertainty by using a sample of remote SNe~Ia.  The
determination of galaxy peculiar velocities is, of course, an
inherently statistical problem because the goal is not to measure a
single number but to measure enough peculiar velocities to map cosmic
flows.

What we must do, then, is to use one or more
distance-independent-observables (DIOs) either (1) to define a subset
of SNe~Ia whose absolute magnitudes are sufficiently homogeneous
(``nearly standard candles"), or (2) to correct the absolute
magnitudes of all or of a subset of SNe~Ia to {\sl make} them
sufficiently homogeneous (``standardized candles").  An already well
known example of the second procedure is to use a light-curve decline
rate, or shape, or width, as the DIO.  This is just the same as using
the periods of Cepheid variables as DIOs to standardize their absolute
magnitudes.

In section 2, we consider the potential use of various photometric and
spectroscopic SN~Ia DIOs, as well as a parent-galaxy DIO.
Absolute-magnitude calibrations and the value of $H_0$ are discussed
in section 3. A few comments concerning the use of SNe~Ia for
measuring $q_0$ are made in section 4, and some summary comments
appear in section 5.

\section{SNe~Ia AS NEARLY STANDARD, OR STANDARDIZED, CANDLES}

In this section we consider ways to define subsets of nearly standard
candles, or of standardized candles.  The approach here is purely
empirical, with matters of physical interpretation being deferred to
section 3.

\subsection{The SN~Ia Absolute-Magnitude Distribution}

The distributions of the blue and visual absolute magnitudes of SNe~Ia
in the current observational sample have been studied by Vaughan et
al. (1995a, 1995b), using Tully-Fisher, SBF, and Hubble-law distances
(the latter with $H_0=85$ to be consistent with the TF and SBF
distance scales).  The distributions are strongly peaked near
``normal" or ``ridge-line" absolute magnitudes (ridge line refers to
plots of magnitude versus distance) of $M_B \simeq M_V \simeq -18.6 -
5 log (H_0/85)$.  In addition to the ridge line SNe~Ia, there are some
SNe~Ia that are dimmer because they are extinguished by dust in their
parent galaxies, and some that are intrinsically dim.  Vaughan et
al. (1995a) showed that it is possible to make an objective definition
of the ``ridge line'', based on the mode of the distribution, that
does not appeal to any DIO.  (The remarks of Hamuy et al. (1995a)
concerning the non-objectivity of this definition reveal a misreading
of the definition.) This mode-based definition was found to isolate a
ridge-line subsample of SNe~Ia that has observational dispersions
$\sigma_{obs}(M_B) \simeq \sigma_{obs}(M_V) < 0.4$ mag; considering
the errors, the intrinsic dispersions, $\sigma_{int}(M_B)$ and
$\sigma_{int}(M_V)$, were judged to be less than 0.2 mag.  Vaughan et
al. (1995b) incorporate the new data of Hamuy et al. (1995a), make
more stringent criteria for inclusion in the sample, and define a
ridge line in terms of a ``binless mode'', to find dispersions
$\sigma_{obs}(M_B) = \sigma_{obs}(M_V) = 0.28$ mag; again the
intrinsic dispersions are judged to be no more than 0.2 mag.  Thus,
{\sl ridge-line SNe~Ia are nearly standard candles.}

An important feature of the SN~Ia absolute-magnitude distribution is
that it declines steeply on its bright side (Vaughan et al. 1995b).
So far, only the spectroscopically peculiar SN 1991T, after correction
for extinction in its parent galaxy, was a substantially overluminous
event.  This steepness on the bright side of the luminosity function
means that the effects of observational selection bias in favor of
bright events at large distances will not be severe.  Still, to the
extent that overluminous SNe~Ia do exist, and that observational bias
does occur, SNe~Ia that get into remote samples may tend to be more
luminous than nearer ones, in which case using the ridge-line for
direct matching of the two samples will tend to underestimate the
distances of the remote events and overestimate the values of both
$H_0$ and $q_0$.  Thus we turn our attention to the use of DIOs to
correct for whatever systematic differences there may be between
remote and nearer SNe~Ia.

\subsection{Photometric SN~Ia DIOs}

Vaughan et al. (1995a,b) use the B-V color at maximum light as a DIO
to define a subset of nearly standard candles.  B-V is available for
most of the well observed SNe~Ia, and it has the attractive feature of
automatically excluding those that are substantially extinguished.
Vaughan et al. (1995a) used the criterion $|B-V| < 0.25$ to define a
subset of normal-color SNe~Ia having $\sigma_{obs}(M_B) \simeq
\sigma_{obs}(M_V) \simeq 0.3$ mag., and intrinsic dispersions of no
more than 0.2 mag.  For the more stringently defined sample of Vaughan
et al. (1995b), the criterion $B-V < 0.2$ is used to define subsets of
normal-color SNe~Ia again having observational dispersions less than
0.3 mag, and intrinsic dispersions less than 0.2 mag.  Thus, {\sl SNe
Ia that have normal B--V colors are nearly standard candles.}

\begin{figure*}
\leavevmode
\psfig{file=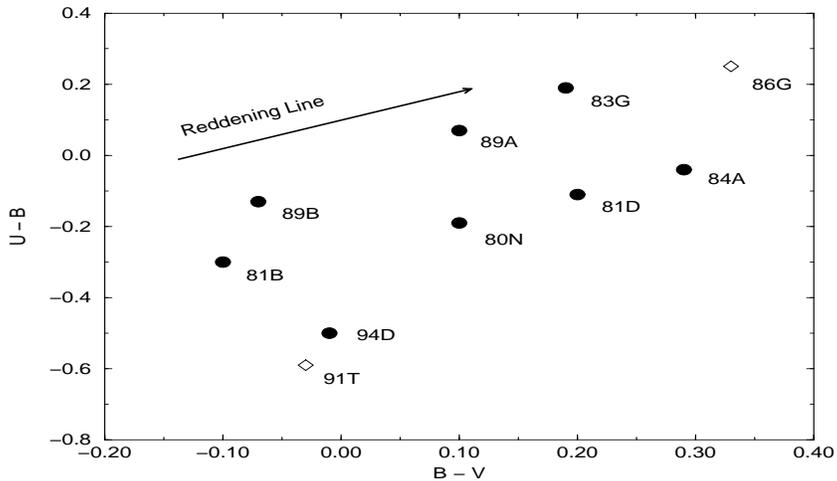,height=3.0in,width=5.0in,angle=270}
\caption{A graph of $U-B$ vs. $B-V$. Open symbols correspond to
spectroscopically peculiar SNe~Ia.\label{ub1}}
\end{figure*}

\begin{figure*}
\leavevmode
\psfig{file=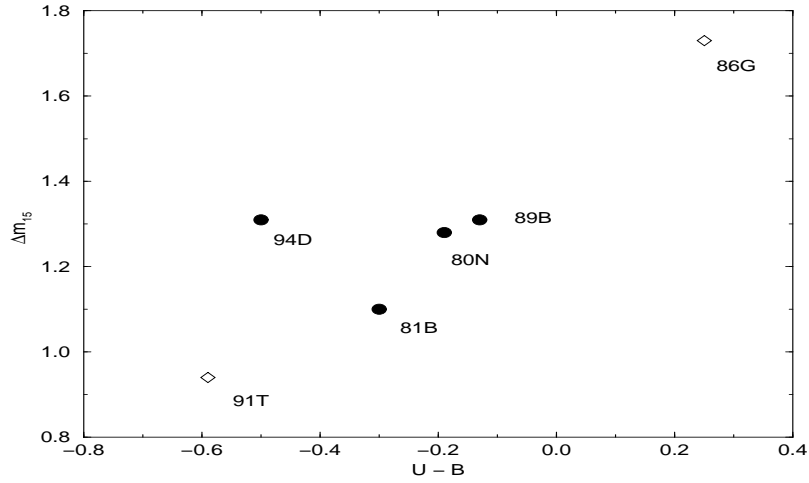,height=3.0in,width=5.0in,angle=270}
\caption{A graph of $\Delta m_{15}(B)$ vs. $U-B$. Open symbols
correspond to spectroscopically peculiar SNe~Ia.\label{dm15ub}}
\end{figure*}

\begin{figure*}
\leavevmode
\psfig{file=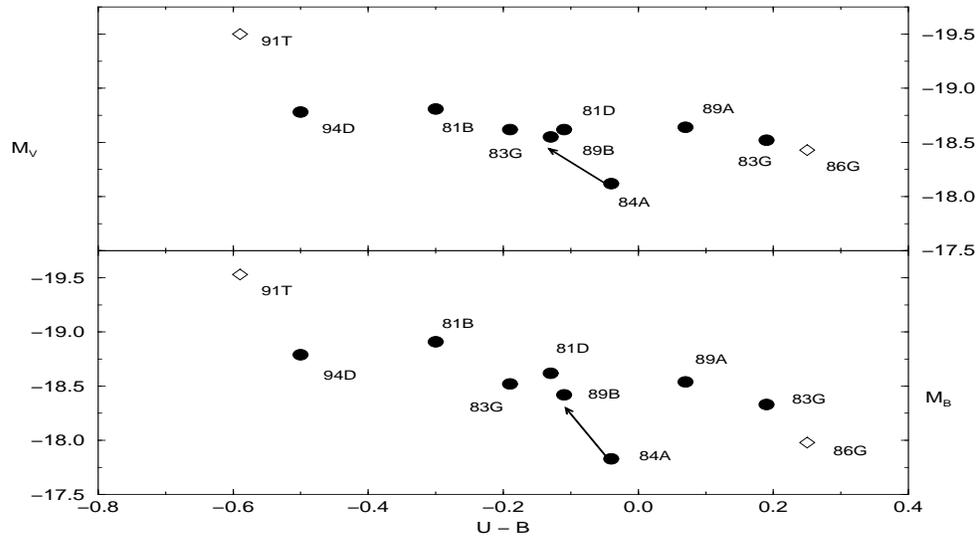,height=3.0in,width=5.0in,angle=270}
\caption{Top panel - $M_V$ vs. $U-B$. Bottom panel - $M_B$
vs. $U-B$. Open symbols correspond to spectroscopically peculiar
SNe~Ia.\label{mbub}} 
\end{figure*}

The $U-B$ color at maximum light also automatically identifies highly
extinguished SNe~Ia, and it is likely to be a more sensitive indicator
of absolute magnitude than is $B-V$.  Schaefer (1995a) critically
discusses what little $U-B$ data is available so far, and presents a
$U-B$ versus $B-V$ diagram.  The one presented here as Fig.~\ref{ub1} is
restricted to only the ``good'' SN~Ia data assembled by Schaefer, plus
a recent measurement for SN 1994D (Patat et al. 1995; Meikle et
al. 1995; Richmond et al. 1995).  In this and subsequent figures, SNe
1986G, 1981B, 1989B, and 1991T have been corrected for the {\sl total}
extinction as estimated by Hamuy et al. (1995b), while the others have
been corrected only for foreground extinction following Burstein \&
Heiles (1988).  A few of the latter, such as SN 1984A, also may have
been substantially extinguished in their parent galaxies.  In Fig.~\ref{ub1},
most of the SNe~Ia can be regarded to form a sequence parallel to the
reddening line.  A reason to think that this is mainly a sequence of
intrinsic SN~Ia properties, rather than just due to extinction, is
found in Fig.~\ref{dm15ub}, which shows that with the exception of SN 1994D,
$U-B$ correlates well with $\Delta m_{15}(B)$, which is not sensitive
to extinction.  Note that the spectroscopically peculiar SN 1991T had
a very negative $U-B$, i.e., it was anomalously bright in the $U$
band.  The other event having a very negative value of $U-B$ is SN
1994D, which was a spectroscopically normal SN~Ia on the system of
Branch, Fisher, \& Nugent (1993), but overluminous for its relatively
rapid decline rate (Richmond et al. 1995; Patat et al. 1995).  A
reason to think that the overluminosity of SN 1994D for its decline
rate is genuine, rather than due to an erroneously long (relative) SBF
distance for its parent galaxy, is that its (distance-independent)
$U-B$ was so unusual.  $M_B$ and $M_V$ are plotted against $U-B$ in
Fig.~\ref{mbub}.  Because $U-B$ appears to correlate well with absolute
magnitude, it may prove to be useful for defining nearly standard
candles or even standardized candles.  The $U$ magnitude is, however,
awkward to measure from the ground.  The main importance of $U-B$ may
be for high-redshift SNe~Ia, where measurements at longer wavelengths
can probe the rest frame wavelengths of the $U$ band.

The other photometric DIOs that have been used recently are based on
the light--curve shape: the $\Delta m_{15}(B)$ parameter of Phillips
(1993) and Hamuy et al. (1995a); the light-curve-shape (LCS) analysis
of Riess, Press, and Kirshner (1995a); or the LBL light--curve stretch
factor (Perlmutter et al. 1995b). Both the $\Delta m_{15}(B)$ and LCS
approaches are said to be capable of standardizing absolute magnitudes
to within dispersions of about 0.1 mag.  As mentioned above, the well
observed SN 1994D appears not to conform to the expectations of the
$\Delta m_{15}(B)$ and LCS techniques, having been too bright for its
light-curve decline rate.  Light-curve-shape DIOs necessarily
require data at phases other than maximum light, and neither the
$\Delta m_{15}(B)$ approach nor the original version of LCS
automatically handle highly extinguished supernovae.  The LCS approach
is now being extended to include $B-V$ colors to solve simultaneously
for the extinction (Riess 1995).

\subsection{Spectroscopic SN~Ia DIOs}

A simple spectroscopic DIO would be ``normalcy'', as defined, for
example, by Branch et al. (1993). Both intrinsically dim events such
as SNe 1991bg, 1986G, and 1992K, and the overluminous SN 1991T, are
{\sl not} spectroscopically normal.  An advantage of this DIO is that
the spectrum does not need to be obtained at any particular phase, as
long as it is early enough to exclude SNe~Ia like 1991T, i.e., not
later than about two weeks after maximum light.  A potential
disadvantage is the non-quantitative definition of ``normal" and the
consequent ambiguity for intermediate cases, if such exist.

The velocity of the red edge of the blueshifted Ca~II H\&K absorption
blend in moderately late time spectra, $V_R(Ca)$, correlates with
absolute magnitude (Fisher et al. 1995).  Because of the limited
number of events for which $V_R(Ca)$ is available, it is not yet clear
whether $V_R(Ca)$ should be used only to define nearly standard candle
subsets, or whether it can also be used to provide standardized
candles.  A disadvantage of $V_R(Ca)$, especially for high-redshift
SNe Ia, is that it is based on observations more than 50 days after
maximum light.  The main importance of $V_R(Ca)$ probably has to do
with its implications for the explosion physics rather than for
distance determinations.

Wells et al. (1994) showed, for a small sample of SNe~Ia, that the
blueshift of the absorption minimum of the Ca~II H\&K blend at maximum
light, $V_0(Ca~II)$, correlates with the $\Delta m_{15}(B)$ parameter.
We have verified that, as would then be expected, $V_0(Ca~II)$ also
correlates with absolute magnitude.  Unfortunately, near maximum light
$V_0(Ca~II)$ varies strongly with phase.

Nugent et al. (1995c) show that in maximum-light spectra, the ratio
$R(Si~II)$ of the fractional depths of the Si~II absorption features
near 5800 and 6150 \AA\ , and the ratio $R(Ca~II)$ of the fluxes of
the emission peaks redward and blueward of the Ca~II H\&K absorption
blend, both correlate with absolute magnitude.  Studies of the phase
dependence of $R(Si~II)$ and $R(Ca~II)$ are underway; preliminary
indications are that near maximum light $R(Si~II)$, at least, does not
vary rapidly with phase.

\subsection{A Parent-Galaxy DIO}

Branch, Romanishin, \& Baron (1995a; here BRB95) argue that there are
convincing connections between the statistical properties of SNe~Ia
and the $B-V$ colors of their parent galaxies.  The blueshift of the
Si~II absorption near 6150 \AA\ in spectra ten days after maximum
light (Branch \& van den Bergh 1993) shows a larger dispersion among
SNe~Ia in red galaxies than among those in blue galaxies, and red
galaxies produce SNe~Ia having both the lowest and the highest Si~II
blueshifts.  The $\Delta m_{15}(B)$ parameter tends to be larger
(faster light-curve decline) in SNe~Ia in red galaxies than in SNe~Ia
in blue galaxies.  And SNe~Ia in red galaxies tend to be dimmer than
SNe~Ia in blue galaxies.  Even when spectroscopically peculiar SNe~Ia,
which include the subluminous SNe 1986G, 1991bg, and 1992K in red
galaxies and the overluminous SN 1991T in a moderately blue galaxy,
are disregarded, the remaining, normal SNe~Ia in red galaxies tend to
be about 0.3 mag dimmer than those in blue galaxies.  The {\sl
observational} dispersions of SNe~Ia in blue and red galaxies,
considered separately, are only $\sigma_{obs}(M_B) \simeq
\sigma_{obs}(M_V)\simeq 0.2$.  Thus, {\sl spectroscopically normal
SNe~Ia in red galaxies and, separately, spectroscopically normal
SNe~Ia in blue galaxies, are very nearly standard candles.}

\section{Absolute-Magnitude Calibration and the Value of $H_0$}

\subsection{Cepheid Calibrations}

Cepheid-based absolute-magnitude calibrations have been reported for
SNe 1937C (Sandage et al. 1992, Saha et al. 1994), SNe 1972E and 1895B
(Sandage et al. 1994; Saha et al. 1995a), and SN 1981B (Saha et
al. 1995b), and preliminary results have been released for SN 1960F
(Sandage 1995, Saha 1995).  Four of these SNe~Ia were
spectroscopically normal (Branch et al. 1993) and the single spectrum
that is available for the fifth, SN 1895B (Schaefer 1995b), is
consistent with normalcy. In addition, as discussed by BRB95, the
Cepheid-based distance to M96 measured by Tanvir et al. (1995) appears
to have effectively measured the distance to a fellow member of the
Leo Spur (Tully 1987), NGC 3627, and its spectroscopically normal SN
1989B. The dispersions of the Cepheid-calibrated absolute magnitudes
of these six SNe~Ia (all in blue galaxies) are very small
($\sigma(M_B)=0.17$, $\sigma(M_V)=0.09$; BRB95).\footnote {NGC 4536,
4496, and 4527, the parent galaxies of SNe 1981B, 1960F, and 1991T,
are thought to be members of the same group of galaxies in the
foreground of the Virgo cluster (e.g., Tully 1988), and the
Tully-Fisher distance moduli to NGC 4536 and 4527 listed by Pierce
(1995), 30.5$\pm$0.3 and 30.6$\pm$0.3, are in agreement within the
uncertainties.  (NGC 4496 is too mildly inclined for a good
Tully-Fisher distance determination.)  Thus, it appears that by
measuring Cepheid-based distances to NGC 4536 and 4496, Saha et
al. have indirectly measured the distance to NGC 4527 and its SN
1991T.  This event was, however, spectroscopically peculiar in a way
that, so far, is singular.  The extreme brightness of SN 1991T (after
correction for extinction) raises interesting questions about its
explosion mechanism and its stellar progenitor, but because SN 1991T
is so unusual we will not use it in connection with $H_0$.  (SN 1991T
{\sl has} entered into several recent determinations of $H_0$, and in
our opinion it has caused some confusion.)}

In the spirit of the previous section, there are a number of ways in
which the Cepheid-calibrated absolute magnitudes can be matched to
those of more remote SNe~Ia in the Hubble flow, to determine the value
of $H_0$: e.g., by means of a straightforward standard-candle
treatment (Tammann \& Sandage 1995; Saha et al. 1995b); a binless mode
or a $B-V$ cut (Vaughan et al. 1995a,b); a correlation between $\Delta
m_{15}(B)$ and absolute magnitude (Hamuy et al. 1995a); an LCS
analysis (Riess et al. 1995a); or $V_R(Ca)$ (Fisher et al. 1995).  All
of these treatments will give nearly the same answer (provided that
the LCS analysis is redone so that it no longer rests on the Phillips
(1993) slope of the light-curve decline-rate correlation, which was too
steep).  Because SNe~Ia in blue galaxies appear to have very small
absolute-magnitude dispersions, and because the Cepheid-calibrated
SNe~Ia are in blue galaxies, at this time we favor the value obtained
by BRB95 by applying the Cepheid calibration to SNe~Ia in blue
galaxies, $H_0=57\pm4$.

A reason to suspect that the actual value of $H_0$ may turn out to be
a little lower is that, as Tammann \& Sandage (1995) emphasize, it is
unacceptable on statistical grounds that the nearby,
Cepheid-calibrated SNe~Ia should be systematically brighter than the
remote SNe~Ia in the Hubble flow.  On the contrary, to the extent that
there is an intrinsic dispersion in SN~Ia absolute magnitudes, the
remote SNe~Ia should be systematically {\sl brighter} than the nearer
ones.  From this kind of reasoning Tammann \& Sandage conclude that an
{\sl upper limit} to $H_0$ is 57 $\pm$8.

\subsection{Physical Calibrations}

There are two simple, instructive ways to think about the peak
luminosity of an SN~Ia in physical terms.  One way is to consider that
the luminosity is the thermal emission from an atmosphere that has a
certain size and temperature structure.  Rather than worry about what
{\sl determines} the properties of the atmosphere, one can use
observed spectra to infer the properties, and then compute the
luminosity.  Early, primitive applications of this approach to SN~Ia
luminosities (Branch \& Patchett 1973; Kirshner \& Kwan 1974; Branch
1979) were referred to as the Baade, or Baade-Wesselink, method.  For
Type II Supernovae, this was succeeded by the more advanced Expanding
Photosphere Method (Eastman \& Kirshner 1989, Schmidt, Kirshner, \&
Eastman 1992).  We have begun to refer to the method, as we (Baron et
al. 1994, 1995: Nugent et al. 1995a,b,c) apply it
to supernovae of all types, as the Spectral-fitting Expanding
Atmospheres Method (SEAM), to emphasize that in our approach one
actually fits a synthetic spectrum to the observed spectra of the
supernova whose distance one is trying to determine, and one makes no
assumptions about the existence of a photosphere. This has been
carried out for SNe~Ia by the program PHOENIX 5.9. This code
accurately solves the fully relativistic radiation transport equation
along with the non-LTE rate equations (for some ions) while ensuring
radiative equilibrium (energy conservation). The following species were
treated in non-LTE: Na~I (3 levels), Ca~II (5 levels), Mg~II (18
levels), O~I (36 levels) and Fe~II (617 levels). These models allow us to
predict a supernova's luminosity given the following assumptions: (1)
Homologous expansion, $v \propto r$, from which it follows that the
velocity of a given matter element is constant and hence that the
position of a reference radius is given by $v_{0}t_{\rm R}$, where
$v_0$ is determined by the spectral fit. (2) The density gradient for
the atmosphere follows a decaying exponential.  (3) A high-quality fit
to the spectrum implies that the model parameters were accurately
determined.

\begin{figure*}
\leavevmode
\psfig{file=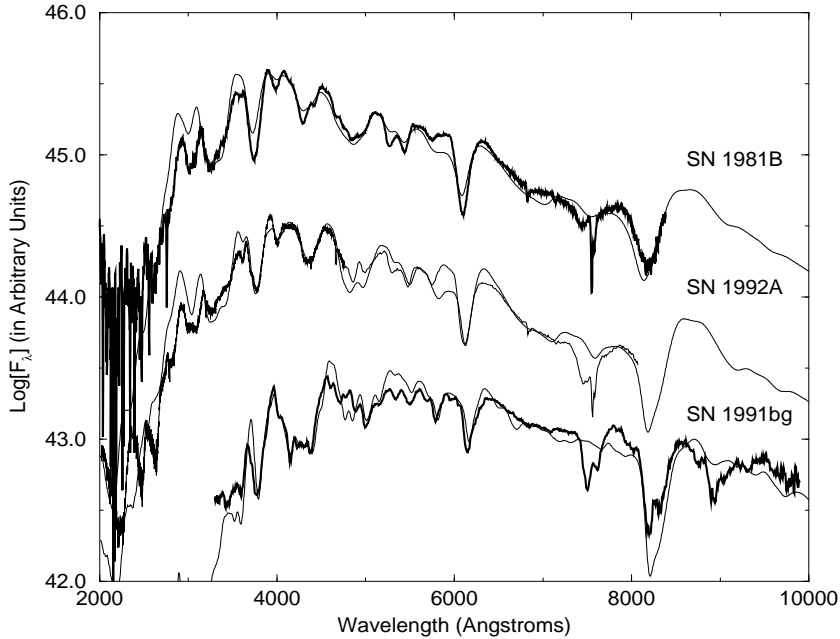,height=4.0in,width=5.0in,angle=270}
\caption{Spectrum synthesis by means of SEAM (thin lines) of SNe
1981B, 1992A and 1991bg (thick lines).\label{snespe}}
\end{figure*}

Fig~\ref{snespe} shows the quality fits obtained for the
near-maximum-light spectra of SNe 1981B (at maximum light), 1992A
(five days post-maximum) and 1991bg (maximum light), and Table 1 lists
the model parameters that were used to produce the synthetic spectra
and the corresponding luminosities and colors. These models were based
on a model--W7 composition in which we homogenized the abundances in
matter moving faster than 8000 km~s$^{-1}$. In the case of SN 1991bg,
we also increased the abundance of Ti~II by a factor of 10.  The rise
times were estimated by means of a procedure similar to the one
described in Nugent et al. (1995a). Two things are readily apparent
from these SEAM results.

\begin{table}[htb]
\begin{center}
\caption{Model parameters for the spectral fits. Velocities are given
in km~s$^{-1}$, temperature in degrees Kelvin and time to peak blue
magnitude in days.}
\begin{tabular}{ccccccccc}
\hline 
SN & $T_{eff}$ & $t$ & $v_{phot}$ & $v_e$ & $M_{bol}$ & $M_{B}$ &
$U-B$ & $B-V$\\ 
\hline
1981B & 9300 & 18 & 11400 & 1500 & -19.38 & -19.49 & -0.48 & -0.07\\
1992A & 8700 & 20 & 8000 & 1400 & -18.65 & -18.73 & -0.15 & 0.12\\
1991bg & 7400 & 15 & 7400 & 800 & -17.16 & -16.67 & 0.58 & 1.00\\
\hline
\end{tabular}
\end{center}
\end{table}

First, these SNe~Ia do indeed form part of the sequence discussed in
Nugent et al. (1995c). There is a clear decline in temperature,
velocity and $v_e$, from SN 1981B to 1992A to 1991bg. The magnitudes
and colors vary, from the blue and bright SN 1981B to the dim and red
SN 1991bg.  Note the over one magnitude range in $U-B$, which agrees
well with observation (see Fig~\ref{ub1}).

Second, given reasonable estimates of the rise times, we obtain
absolute magnitudes and distances. To estimate our errors we
conservatively assume errors of 0.2 mag. from the velocity, 0.2
mag. from the composition, and 0.2 mag. for the fact that not all
species are treated in non-LTE. Adding these in quadrature we arrive
at errors of $\pm$ 0.35 (for the estimated rise times). Table 2
displays the resulting peak absolute magnitudes for each and their
corresponding distance moduli (We have allowed for the fact that at
the epoch of the SN 1992A spectrum, five days post--maximum, the light
curve had declined 0.3 mag. in the blue from its peak (Suntzeff
1992)).

\begin{table}[htb]
\begin{center}
\caption{Peak absolute magnitudes, apparent magnitudes (Nugent et
al., 1995c) and the corresponding distance moduli for the spectral
fits.}     
\begin{tabular}{ccccccc}
\hline 
SN & $M_{B}$ & $M_V$ & $B$ & $V$ & $E_{B-V}$ & $\mu$\\
\hline
1981B & -19.49(35) & -19.42(35) & 12.03(05) & 11.93(05) & 0.10(05) &
31.08(36)  \\
1992A & -19.03(35) & -19.02(35) & 12.57(03) & 12.58(03) & 0.00(02) &
31.60(35)   \\
1991bg & -16.67(35)& -17.67(35) & 14.76(10) & 14.02(05) & 0.00(02) &
31.62(35)  \\
\hline
\end{tabular}
\end{center}
\end{table}

Using the relation between absolute magnitude and H$_0$ given in
section 2.1, the mean SEAM absolute magnitudes of the normal SNe 1981B
and 1992A correspond to $H_0 = 63 \pm 11$, where we have allowed for
an uncertainty of 2 days in the rise times.

\begin{figure*}
\leavevmode
\psfig{file=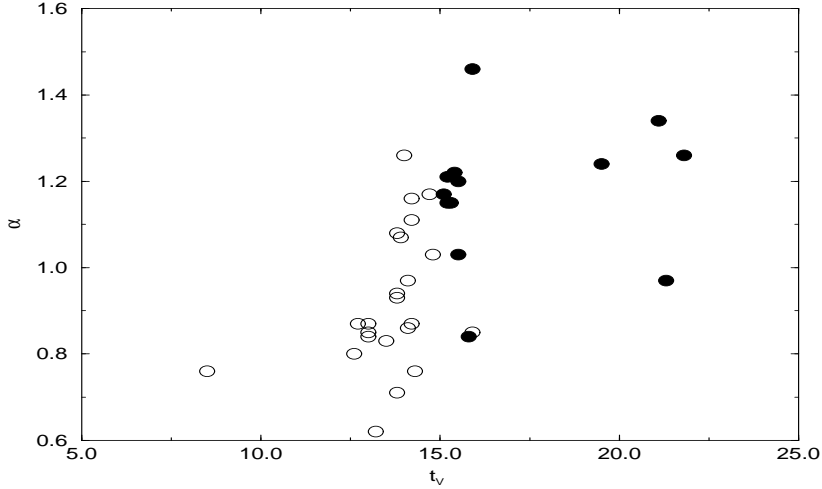,height=3.0in,width=5.0in,angle=270}
\caption{$\alpha$ vs. $t_V$. Open circles correspond to He--ignitor
models and/or models with a rise time less than 15 days.\label{alpha}}
\end{figure*}

\begin{figure*}
\leavevmode
\psfig{file=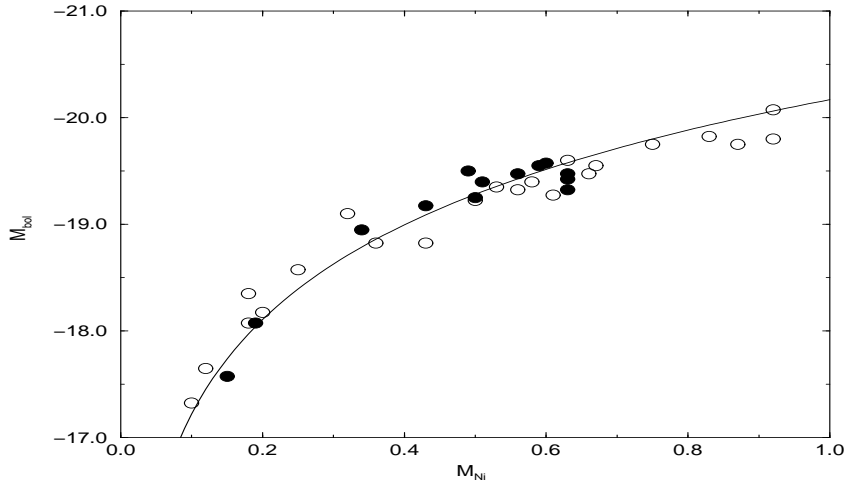,height=3.0in,width=5.0in,angle=270}
\caption{$M_{bol}$ vs. $M_{Ni}$. Open circles correspond to He--ignitor
models and or models with a rise time less than 15 days. The shape of
the line corresponds to the simple case in which the peak luminosity
is strictly proportional to $M_{Ni}$.\label{mni}}
\end{figure*}

\begin{figure*}
\leavevmode
\psfig{file=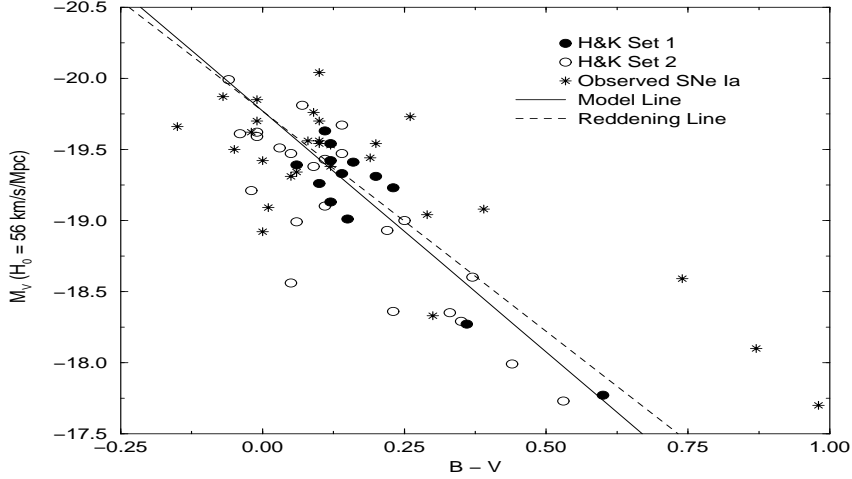,height=3.0in,width=5.0in,angle=270}
\caption{A graph of $M_V$ vs. $B-V$. Open circles correspond to 
He--ignitor models and/or models with a rise time less than 15
days.\label{hnk}}
\end{figure*}

Alternatively, one can think about the peak luminosity as being
powered by the gamma-ray and positron products of the radioactive
decay chain $^{56}$Ni $\rightarrow$ $^{56}$Co $\rightarrow$ $^{56}$Fe.
Then a natural first approximation to the luminosity is:
\begin{equation}
L_{\rm Ni} = \alpha \dot S(t_{R}) M_{\rm Ni}
\end{equation}
where $\dot S$, the instantaneous radioactivity luminosity at the time
of maximum light, is a known function of the rise time, $t_r$, $M(Ni)$
is the ejected mass of $^{56}$Ni, and $\alpha$ is a dimensionless
parameter of order unity.  For certain simplifying assumptions
$\alpha$ {\sl is} unity (Arnett 1982, Branch \& Khokhlov 1995).  More
realistic values of $\alpha$ can be obtained only from detailed
light-curve calculations which incorporate the dependence of the
opacity on the physical conditions and the composition.  Such
calculations have been carried out by Harkness (1991), and by
H\"oflich and his collaborators; the latter are summarized by
H\"oflich \& Khokhlov (1995; here H\&K95).  For the very wide variety
of models considered by H\&K95, the value of $\alpha$ ranges from 0.62
to 1.46.  However, many of the models are not really ``in the
running'' as representations of the real SNe~Ia that have been
observed; e.g., some are ``helium ignitors'' (sub-Chandrasekhar white
dwarfs whose first nuclear ignition is far off-center near the base of
an accreted helium layer, rather than in carbon-oxygen near the center
of a Chandrasekhar-mass white dwarf) which probably cannot satisfy
observed light curves and spectra (H\&K95; Branch et al. 1995b; Nugent
et al. 1995x), and some have rise times to maximum light that are too
short, ($<$ 15 days).  Fig.~\ref{alpha}, in which $\alpha$ is plotted
against the rise time to bolometric maximum light for all of the
H\&K95 models, shows that low values of $\alpha$ tend to be associated
with models that rise to maximum too fast.  (In
Figs.~\ref{alpha},~\ref{mni} and~\ref{hnk}, open circles denote models
that are not really in the running because they are helium ignitors or
they have rise times of less than 15 days.) For models which have rise
times of 15 days or more, the value of $\alpha$ tends to be exceed
unity (for physical reasons that are explained by H\&K95 and
references therein).  If we adopt M(Ni)=0.6$\pm0.1$ M$_\odot$
(Harkness 1991; H\"oflich 1995), $\alpha=1.2 \pm 0.2$, and a rise time
of 18$\pm$2 days as characteristic values for normal SNe Ia, we obtain
$M_{bol}=-19.3 \pm 0.3$.  With a characteristic bolometric correction
for normal SNe Ia of 0.1 mag, we have $M_B \simeq M_V \simeq -19.4 \pm
0.2$, which corresponds to $H_0=59 \pm 9$.

There are several alternatives to the $\alpha$ formulation of the
$^{56}$Ni approach to peak luminosities.  For example, Fig.~\ref{mni}
shows the bolometric absolute magnitudes plotted against $M(Ni)$ for
the models of H\&K95.  This figure is just like one shown by H\&K95,
but here the models in the running as representations of observed SNe
Ia are identified by filled circles.  It is very interesting that in
spite of the fact that the models have a variety of rise times and
$\alpha$ parameters, the peak luminosity turns out to be, to a good
approximation, proportional to the nickel mass.  For the range of
interest for normal SNe~Ia, $M(Ni)=0.6 \pm 0.1$, the bolometric
magnitude is $-19.5\pm0.1$, which with the other parameters as above
leads to $H_0=56\pm8$.

Another way to use the H\&K95 models is discussed recently
by van den Bergh (1995; here vdB95), who plots $M_V$ against $B-V$ for
the models and notes that the trend is nearly degenerate with an
interstellar reddening line.  He points out that consequently one can
define a quantity $M_V=M_*+3.1(B-V)$, and predict $M_V$ for an SN~Ia of
an observed $B-V$ without knowing the reddening.  Fig.~\ref{hnk} is
just like Fig. 1 of vdB95, but with the H\&K95 models that are in the
running being identified by filled circles, and with observed SNe~Ia
(Vaughan et al. 1995b) also plotted, as asterisks.  Note that the
models that are in the running are even less scattered than are the
models in general.  The solid line, a linear fit to the models in the
running, is $M_V=-19.80+3.39(B-V)$; thus its slope is close to the
reddening slope of 3.1.  Adopting the model slope of 3.39 and fitting
all of the real SNe, except the three reddest ones for which there are
no models, to the model line yields $H_0=56\pm5$.  (If the three
reddest SNe~Ia were included the result would be $H_0=59\pm5$).  Thus
we obtain a result that agrees with that of vdB95. Where we differ
with vdB95 is in the choice of conclusions.  Those offered by vdB95
are either (1) the H\&K95 models are too bright; (2) the H\&K models
are too red, or (3) the Cepheid zero point is wrong by 0.75 mag. The
latter conclusion is based on contrasting the low value of $H_0$ that
is obtained from the SN Ia models of H\&K95 to the the higher value of
$H_0$ that has been derived on the basis of the Cepheid--based
distances to galaxies associated with the Virgo cluster (Pierce et
al. 1994; Freedman et al. 1994; Kennicutt et al. 1995; Mould,
Freedman, \& Kennicutt 1995; vdB95).  {\sl However, the low value of
$H_0$ from the H\&K95 models is in excellent agreement with the result
obtained in section 3.1, from Cepheid-based distances to parent
galaxies of SNe~Ia!}

Therefore we offer choice (4): the route to $H_0$ through Virgo
galaxies is a hazardous one, which currently overestimates $H_0$.  Not
so much because there is anything wrong with the Cepheid zero--point,
but in part because of the use of too large of an ``unperturbed cosmic
velocity'' for Virgo (e.g., Freedman et al. 1994 use 1404 km~s$^{-1}$
while Jerjen \& Tammann (1993) find 1179 km~s$^{-1}$ and Gudehus
(1995) finds about 1000 km~s$^{-1}$), and in part because, owing
perhaps to selection effects, the ``Virgo'' spirals whose distances
have been measured on the basis of Cepheids, so far, tend to be on the
near side of the very extended spatial distribution of the Virgo
spirals.  Note that {\sl before} their Cepheid--based distances were
measured, NGC 4536 and NGC 4496 were found to be on the front side by
Tully (1988), from a Virgocentric flow model, and by Pierce (1994),
from a Tully--Fisher analysis. This point of view is supported by a
preliminary analysis of the distance to Virgo by means of SEAM (Nugent
et al. 1996). Table 2 lists the SEAM distances to SNe 1981B in NGC
4536 and 1991bg in NGC 4374, an elliptical galaxy in Virgo. SEAM gives
the same distance modulus to SN 1981B and NGC 4536 as do the Cepheids,
31.1 mag. ($\approx$ 17 Mpc). But the SEAM distance modulus to SN
1991bg and NGC 4374 is 31.6 mag. ($\approx$ 21 Mpc). This difference
of 0.5 mag. is the same as is implied by Tully--Fisher analysis of NGC
4536 and SBF analysis of NGC 4374. Thus if one believes that the
ellipticals are at the center of Virgo, and that SEAM and/or
Tully--Fisher plus SBF provide the correct difference in distance
modulus between NGC 4536 and NGC 4374, then one arrives at a distance
to Virgo of 21 Mpc. Then the value of $H_0$ obtained via the Virgo
cluster is easily reconciled with the low value obtained above from
SNe Ia in general: for example, a plausible mean distance to Virgo of
21 Mpc, and a plausible unperturbed velocity of 1100 km~s$^{-1}$,
would give $H_0 = 52$ from Virgo.

SEAM and the $^{56}$Ni method are not really separate, of course.  The
amount of ejected $^{56}$Ni determines the atmospheric structure, and
the two approaches are destined to merge into one.  Nugent et
al. (1995b) combined the two approaches, but only to eliminate the
rise time dependence, which enters in opposite directions when these
two approaches are considered separately.  Combining the two in a more
fundamental sense has been begun by Harkness (1991) and H\"oflich
(1995) who attempt to simultaneously match both light curves and
spectra with a given model.

\begin{figure*}
\leavevmode
\psfig{file=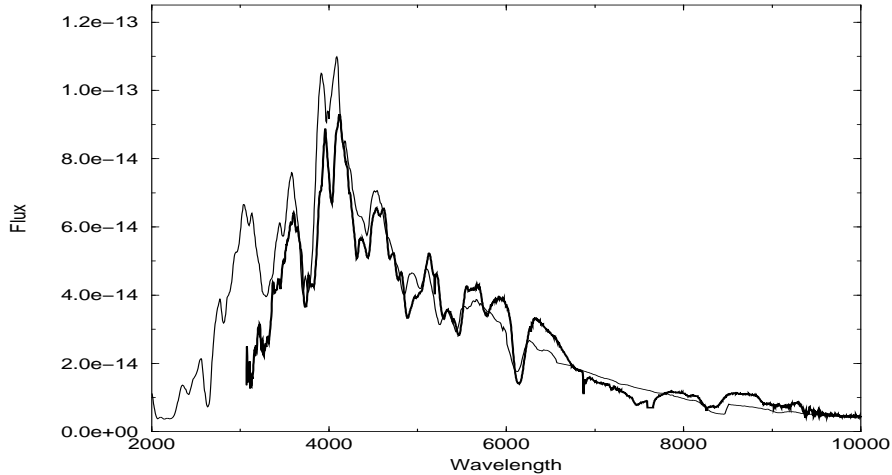,height=3.0in,width=5.0in,angle=270}
\caption{A comparison of the spectrum of H\"oflich's (1995) model M36 
(thin line) vs. the maximum light spectrum of SN 1994D.\label{pah}
Note the discrepancy near 3000 \AA .}
\end{figure*}

As in the previous section on $H_0$ from Cepheid-calibrated SNe~Ia, we
conclude this section with a reason to think that the true value of
$H_0$ may be a little lower than obtained here from physical
considerations.  In Fig.~\ref{pah}, H\"oflich's (1995) calculated
spectrum of model M36 is compared with the observed spectrum of SN
1994D (Meikle et al. 1996).  The flux excess in the synthetic spectrum
near 3000 \AA\ suggests an insufficiency of ultraviolet line blocking
in the calculations.  With more line blocking, the excess ultraviolet
flux would be redistributed to longer wavelengths, where it would make
$M_B$ and $M_V$ brighter.  SEAM also could be underestimating the SN
Ia luminosities if we are underestimating the rise times.

\section{SOME $q_0$ ISSUES}

Already about ten SNe~Ia in the redshift range 0.3 $\simeq z \simeq$
0.5 have been discovered (Norgaard--Nielsen et al. 1989; Perlmutter et
al. 1994, 1995a, 1995b, Schmidt et al. 1995).  Preliminary indications
are that when the eight remote SNe~Ia found by the LBL group are
matched to nearer SNe~Ia on the basis of light-curve width --- using the
LBL ``stretch factor'' in lieu of $\Delta m_{15}(B)$ or LCS --- the
value of $q_0$ comes out to be near 0.5 (Perlmutter et al. 1995b).

In principle, concerns about evolution with cosmic epoch of the SN Ia
population and about observational selection effects (in favor of the
brightest SNe Ia, or in favor some kind of parent galaxy), are
alleviated by the use of the LBL light--curve stretch factor as a DIO
to standardize the SNe~Ia.  SN 1994D warns us, however, that
luminosity does not correlate perfectly with light--curve shape.  If
the rapidly declining SNe~Ia in the LBL high-redshift sample happened
to be like SN 1994D, then the stretch factor would overcorrect the
luminosities of those SNe~Ia, make them too bright, and {\sl
over}estimate the value of $q_0$.  For this reason, and also because
some high--redshift SNe Ia may not be well observed at phases other
than maximum light, it will be important for a definitive
determination of $q_0$ to develop and refine some of the other DIOs
that we discussed in section 2.

To the extent that the cosmic matter distribution is not homogeneous,
the determination of $q_0$ from the Hubble diagram in the standard way
can lead to an {\sl under}estimate of $q_0$ (Kantowski, Vaughan, \&
Branch 1995).  It is tempting to speculate about what the matter
distribution might be like, and what correction might be required, for
any measured value of $q_0$.  If $q_0$ were measured to be small, say
0.1, then we probably could conclude that the matter in the universe
is clumped, in the sense of Kantowski et al.  Then the measured value
$q_0=0.1$ perhaps should be corrected to $q_0=0.15$, which would be
quantitatively, but not qualitatively, significant.  On the other
hand, if $q_0$ is measured to be high, near 0.5, then the matter
distribution probably is rather smooth, and no correction may be
required.

\section{SOME COSMIC IMPLICATIONS}

SNe Ia, whether calibrated by Cepheids in their parent galaxies, by
spectrum fitting (SEAM), or by some version of the $^{56}$Ni method,
indicate that the value of the Hubble constant is low, less than or
about 60.  (We see no prospect of reconciling SNe Ia with a high value
of $H_0$, such as 80.)  Preliminary indications from the high-redshift
SNe Ia discovered by the LBL group are that $q_0$ is near 0.5
(Perlmutter 1995b).  It is interesting to recall that these are the
values of $H_0$ and $q_0$ that were advocated by Fowler (1987, 1989),
who wanted to close the universe with baryonic matter.  They also are
the values that are required by the decaying--neutrino theory of
Sciama (1990, 1994), if the universe is to be closed by such
neutrinos.

But, for values like these, is there a cosmic age problem?  That
depends to a great deal on how literally one takes the current
estimates of $H_0$, the globular cluster ages, and $q_0$.  We are
tempted to take the point of view that if $q_0$ is being measured to
be {\sl near} 0.5, it probably {\sl is} 0.5, because 0.5 is such a
special number.  Then $H_0=55$, for example, gives an expansion age of
12 Gyr, which according to some is an uncomfortably short amount of
time for globular clusters and for nucleocosmochronology.  But
introducing a non--zero value of $\Lambda$, at 120 powers of ten from
its ``natural'' value (to the extent that $\Lambda$ does have a
natural value), to resolve whatever mild discrepancy there may be
between current estimates of the cosmic expansion timescale and other
timescales, would be incredibly fine tuning.  $H_0$ and the globular
cluster age, unlike q$_0$, have no special values, so we need to
respect their error bars.  In view of the grand lesson of history -
that people get their $H_0$'s too high - and in view of the tendency
for recent estimates of globular cluster ages (e.g., Charboyer et
al. 1995; Shi 1995; Mazzitelli, D'Antona, \& Caloi 1995) to be
somewhat lower than many previous estimates (but not Fowler's), we
conjecture that fine tuning by means of a non--zero $\Lambda$ will not
be required.  That is, {\sl even with $q_0=0.5$, there probably is no
cosmic age problem.}

\section*{Acknowledgments}
We would like to thank the organizers, Ramon Canal, Pilar
Ruiz--Lapuente, and Jordi Isern, for making this meeting possible, and
all the participants for making it so stimulating.  We also thank our
esteemed SEAM collaborators, Eddie Baron and Peter Hauschildt, for
allowing us to present preliminary results for SN 1991bg.  Our work
has been supported by NSF and NASA.

\end{document}